# Nonlinear Dynamics of Composite Fermions in Nanostructures

R. Fleischmann, T. Geisel, C. Holzknecht, and R. Ketzmerick

*Institut für Theoretische Physik und SFB Nichtlineare Dynamik,*
*Universität Frankfurt, D-60054 Frankfurt/Main, Germany*

We outline a theory describing the quasi-classical dynamics of composite fermions in the fractional quantum Hall regime in the potentials of arbitrary nanostructures. By an appropriate parametrization of time we show that their trajectories are independent of their mass and dispersion. This allows to study the dynamics in terms of an effective Hamiltonian although the actual dispersion is as yet unknown. The applicability of the theory is verified in the case of antidot arrays where it explains details of magnetoresistance measurements and thus confirms the existence of these quasi-particles.

Considerable progress in the understanding of interacting 2D electron systems was made when the fractional quantum Hall effect [1] was explained as the integer quantum Hall effect of novel particles, so-called *composite fermions* (CFs) [2]. They result from a singular gauge transformation [3] and consist of an electron with an even number of magnetic flux quanta attached to it. In a mean field approximation they experience an effective magnetic field $B_{\rm eff} = B - 2m\phi_0 n$ given by the external magnetic field $B$ plus the average magnetic field $-2m\phi_0 n$ bound to the other CFs, where $2m$ is the number of magnetic flux quanta $\phi_0 = h/e$ per CF and $n$ is the 2D electron density. At first, CFs seemed to be no more than a helpful mathematical construct. The situation changed, when in a key experiment by Kang et al. [4] it was verified that near filling factors $\nu = 1/2m$ there is a well-defined Fermi surface as predicted by Halperin, Lee, and Read [5], on which CFs seem to move in a similar way as electrons do near $B = 0$. By measuring the magnetoresistance in antidot arrays they found commensurability peaks due to CFs at small *effective* magnetic fields $B_{\rm eff} \simeq 0$, i.e. $B \simeq B_{\nu=1/2m} = 2m\phi_0 n$, that had previously been observed for electrons at small magnetic fields $B \simeq 0$ [6,7] and were explained by classical chaotic electron dynamics in a Sinai billiard with soft walls [8]. Similarly, Goldman et al. [9] observed magnetic focussing peaks of CFs corresponding to an effect that is well understood for electrons [10].

Although the electron analogy in these experiments indicated the existence of CFs, the magnetoresistance near $B_{\rm eff} \simeq 0$, however, conspicuously differed from the magnetoresistance of electrons near $B \simeq 0$. E.g. the antidot experiment showed asymmetries around effective magnetic field zero that are not present for electrons [4] and the magnetic focussing peaks, surprisingly, were stronger for CFs than for electrons [9]. These differences were expected to be related to the striking property of CFs that a nanostructure potential $U(x,y)$, which influences the electron density $n(x,y)$ causes a spatially modulated effective magnetic field [5,11]

$$B_{\rm eff}(x,y) = B - 2m\phi_0 n(x,y). \qquad (1)$$

Thus the motion of CFs at $B \simeq B_{\nu=1/2m}$ should be quite distinct from the motion of electrons at $B \simeq 0$ as soon as the potential is not uniform, which is the case in any experimentally realized nanostructure due to the softness of the confining walls. Quasi-classical equations of motion describing CFs in nanostructure potentials, however, have not been obtained so far thus preventing a detailed analysis of any transport experiment on CFs (e.g. Refs. [4,9,12]).

In the present letter we outline a quasi-classical theory for the motion of CFs in general nanostructure potentials $U(x,y)$ leading to a description in terms of effective potentials $U_{\rm eff}(x,y)$ experienced by the CFs. By an appropriate nonlinear parametrization of time we obtain an effective Hamiltonian with the amazing property that the CF trajectories are independent of the CF mass and more generally of the CF dispersion $\varepsilon_{\rm CF}(\mathbf{k})$. This is particularly important as it allows to study the dynamics of CFs even if we lack the dispersion relation, which as yet is unknown [5,13]. On the other hand we can conclude that an experimental determination of the CF mass and dispersion in the quasi-classical regime will be feasible only by means of time or frequency dependent methods, as only the time scale of the trajectories depends on the dispersion. This theory provides the framework for studying the quasi-classical dynamics and magnetotransport of CFs in any nanostructure as well as its differences from the electron case. We test its applicability in the example of antidot arrays where we calculate the longitudinal and the Hall resistance near filling factor $\nu = 1/2$ and find a surprisingly good agreement with the experiment of Kang et al. [4]. The origins of various experimental features such as a suppression of peaks and asymmetries with respect to an inversion of $B_{\rm eff}$ are explained. The detailed agreement of experiment and theory thus empirically confirms the existence of CFs.

In analogy to electrons [14] the motion of CF wavepackets is described by the *quasi-classical equations of motion* (QCEM) $\mathbf{v} = (1/\hbar)\partial\varepsilon_{\rm CF}(\mathbf{k})/\partial\mathbf{k}$ and $\hbar\dot{\mathbf{k}} =$

$-e(\boldsymbol{v} \times \boldsymbol{B}_{\text{eff}}) - \partial U_{\text{eff}}/\partial \boldsymbol{r}$ for a dispersion $\varepsilon_{\text{CF}}(\boldsymbol{k})$ in a slowly varying potential $U_{\text{eff}}$ (see below) and small effective magnetic field $\boldsymbol{B}_{\text{eff}}$. We expect the QCEM to be a good approximation as long as the Fermi wavelength $\lambda_F$ is small compared to the spatial scale of the potential, as is the case e.g. in antidot arrays [4] ($\lambda_F \simeq 45nm \ll a \simeq 500nm$). Naively, one would expect that CFs experience the same potential $U(x,y)$ as electrons. This leads to the following contradiction, however: As the dispersion $\varepsilon_{\text{CF}}(\boldsymbol{k})$ originates from electron-electron interaction, it is distinct from the electron dispersion [5,15,13]. Therefore CFs and electrons have different Fermi energies, which would lead to very different density distributions if they experienced the same potential. In contrast, for potentials slowly varying on the scale of the magnetic length and for $\nu(x,y) \approx 1/2$ away from any Hall plateaus, it is reasonable to assume that the CF density distribution is the same at $B_{\text{eff}} \simeq 0$ as the electron density $n(x,y)$ at $B \simeq 0$, which is related implicitly to the electrostatic potential by means of the constant electronic Fermi energy $E_F = \varepsilon(n(x,y)) + U(x,y)$. This assumption and the constancy of the CF Fermi energy can only be fulfilled if the CFs experience an effective potential $U_{\text{eff}}$, different from the bare potential $U$, given by

$$U_{\text{eff}}(x,y) = E_F^{\text{CF}} - \varepsilon_{\text{CF}}(n(x,y)). \qquad (2)$$

Inserting this effective potential into the QCEM and introducing a nonlinear parametrization $s(t)$ of time by $ds/dt = (1/\alpha)(d\varepsilon_{\text{CF}}/dn)|_{n=n(x(t),y(t))}$ yields

$$\frac{d\boldsymbol{r}}{ds} = \frac{\alpha}{\hbar}\frac{\partial n}{\partial \boldsymbol{k}}; \qquad \hbar\frac{d\boldsymbol{k}}{ds} = -e\left(\frac{d\boldsymbol{r}}{ds}\times \boldsymbol{B}_{\text{eff}}\right) + \alpha\frac{\partial n}{\partial \boldsymbol{r}}. \qquad (3)$$

Here $\alpha$ is an arbitrary constant giving $s$ the dimension of time, and $n = |\boldsymbol{k}|^2/4\pi$ [5] holds for the spin polarized CFs. Amazingly, these equations of motion are independent of the CF energy dispersion $\varepsilon_{\text{CF}}(\boldsymbol{k})$. Thus $\varepsilon_{\text{CF}}(\boldsymbol{k})$ changes the time scale, but *not* the path of a classical CF trajectory. This property is of great importance, as it allows the study of classical dynamics of CFs even though their dispersion is so far unkown [5,13].

In order to compare with previous work on electron dynamics in nanostructures we derive an equivalent Hamiltonian from Eq. (3): It is convenient to choose $\alpha = \pi\hbar^2/m_e^*$, which with $n(x,y) = (m_e^*/\pi\hbar^2)(E_F - U(x,y))$ and the effective electron mass $m_e^*$ yields

$$H = \frac{(\boldsymbol{p} - e\boldsymbol{A}_{\text{eff}}(x,y))^2}{2(2m_e^*)} + U(x,y) \qquad (4)$$

and allows to describe CFs as if they had the Fermi energy $E_F$ and twice the mass of the electrons and moved in the *bare* potential $U(x,y)$. The factor of 2 originates from the spin polarization of CFs used in Eq. (3). It should be noted, however, that the time or energy scale of Eq. (4) is arbitrary (as was $\alpha$) and should not be confused with the true but unknown time or energy scale of

CFs. The vector potential $\boldsymbol{A}_{\text{eff}}(x,y)$ is determined such that rot $\boldsymbol{A}_{\text{eff}} = B_{\text{eff}}\mathbf{e}_z$ with

$$B_{\text{eff}}(x,y) = B_{\text{eff}}^0 + B_{\nu=\frac{1}{2}}\frac{U(x,y)}{E_F}, \qquad (5)$$

as follows from Eq. (1) for $m=1$. Here $B_{\text{eff}}^0 = B - B_{\nu=1/2}$ is the effective magnetic field for zero potential.

Furthermore, we have to consider that a current of CFs is also a current of magnetic flux quanta and by Faraday's law induces a perpendicular electric field. This effect is important for *non-equilibrium* currents [16] and can easily be incorporated by adding $2mh/e^2$ to the Hall resistance $\rho_{xy}$ calculated in the CF picture. *Equilibrium* currents induced by an electrostatic potential (e.g. diamagnetic currents) would in addition require a selfconsistent determination of the CF dynamics and the effective potential. In the classical limit, however, the equilibrium currents vanish for electrons and CFs [17].

The frequency dependent conductivities $\sigma_{ij}(\omega)$ are given by the classical Kubo formula [8]. For $\omega = 0$ they are invariant under the transformation $t \to s$ [17], allowing us to study DC-magnetotransport of CFs in the classical limit without knowing the CF energy dispersion. From this one can conclude that an experimental determination of the CF mass and dispersion in the quasiclassical regime will be feasible only by means of time or frequency dependent methods.

We now apply these results to the CF dynamics in antidot lattices near $\nu = 1/2$, but first quote some of the main results for small magnetic fields [6–8,18,19]. The electron resistivity $\rho_{xx}(B)$ exhibits a characteristic series of peaks [6,7]. These peaks were shown to be caused by the trapping of chaotic electron trajectories close to nonlinear resonances that reflect commensurabilities of the cyclotron diameter $d_c$ and the lattice period $a$ and correspond to cyclotron-like motion around a certain number of antidots [8]. The number of peaks, their shape and magnetic field positions depend on the parameters of the potential $U(x,y) = U_0[\cos(\pi x/a)\cos(\pi y/a)]^\beta$ [8,19]. For the previously used parameters $\beta = 4$ and $d/a = 1/4$ [8] the electron resistivity $\rho_{xx}$ (Fig. 1a) exhibits two peaks for positive and negative magnetic fields, respectively, and agrees reasonably well with the experimental electron resistivity measured by Kang et al. (Fig. 2 in Ref. [4]) for $a = 700nm$. We therefore use these parameters of the potential for studying the magnetotransport of CFs.

We numerically evaluate the Kubo formula for the CF dynamics using the methods developed for electrons in Ref. [8]. Like for electrons the resistivities are mainly determined by chaotic trajectories (which stay close to the bottom of the antidot potential thus fulfilling the assumption $\nu(x,y) \approx 1/2$). Figure 1a shows the results of these calculations for CFs in the vicinity of $\nu = 1/2$, i.e. $B_{\text{eff}}^0 \simeq 0$. In contrast to the electron result they show (i) a strong asymmetry in the heights of the main peaks for negative and positive effective magnetic fields, (ii) a shift in the peak position at negative effective magnetic fields, (iii) a broadening of their width, and (iv) a supression of the peaks due to CF trajectories encircling 4 antidots. As we have used the same mean free path for CFs and electrons in Fig. 1a, this supression is caused by the spatially varying magnetic field (Eq. (5)), which strongly restricts the possibility of cyclotron-like motion around 4 antidots. In the experiment the mean free path of CFs is considerably smaller than for electrons and therefore these peaks vanish completely. Considering the lack of parameter fitting these results (i)-(iv) agree surprisingly well with the experimental findings by Kang et al. [4], as shown in Fig. 1b for a realistic mean free path.

The most striking feature is the occurrence of an asymmetry in $\rho_{xx}$ around $B_{\text{eff}}^0 = 0$ that is not present for electrons. Why does a reversal of the direction of $B_{\text{eff}}^0$ change the transport properties? From Eq. (5) it is clear that whenever the potential increases, the effective magnetic field $B_{\text{eff}}(x,y)$ experienced by a CF increases. For the case $B_{\text{eff}}^0 > 0$ this will increase the magnitude $|B_{\text{eff}}(x,y)|$, whereas for $B_{\text{eff}}^0 < 0$ it will decrease the magnitude $|B_{\text{eff}}(x,y)|$ and for strong potentials like the antidot potential $B_{\text{eff}}(x,y)$ will even become positive. Thus a sign reversal of $B_{\text{eff}}^0$ does not always lead to a sign reversal of $B_{\text{eff}}(x,y)$ and quite different classical trajectories with flipping curvature can be expected, as shown in Fig. 2a,b (left insets) for trajectories close to a single antidot.

In Fig. 2 the consequences of this asymmetry are shown on a larger length scale for positive and negative effective magnetic fields $B_{\text{eff}}^0$ corresponding to the main peak positions of $\rho_{xx}$. Even though both chaotic trajectories encircle single antidots for some time, their overall behaviour is quite distinct due to curvature flipping for $B_{\text{eff}}^0 < 0$. This explains the existence of the magnetotransport asymmetry around $B_{\text{eff}}^0 = 0$. For various antidot potentials with different antidot sizes and shapes we have found the main peak in $\rho_{xx}$ to be larger for $B_{\text{eff}}^0 < 0$ than for $B_{\text{eff}}^0 > 0$, in agreement with the experimental findings [4].

The shift of the main peak for *negative* magnetic fields (ii) reflects the fact, that the peaks do not follow from simple commensurability arguments, but stem from non-linear resonances in phase space due to complex nonlinear dynamics, as illustrated in the right inset of Fig. 2a. Similar shifts of peak positions have been observed for electrons [6,8,19].

In the future it will be of interest to see how far the insensitivity of the dynamics and the magnetoresistance to the actual dispersion carries over into the regime farther away from $\nu = 1/2m$ where a quantum description is necessary.

We are grateful to H. L. Störmer for extensive stimulating discussions on the slopes of Mauterndorf and acknowledge further discussions with J. K. Jain, G. Kirczenow, N. Read, and D. Weiss. This work was supported by the Deutsche Forschungsgemeinschaft.

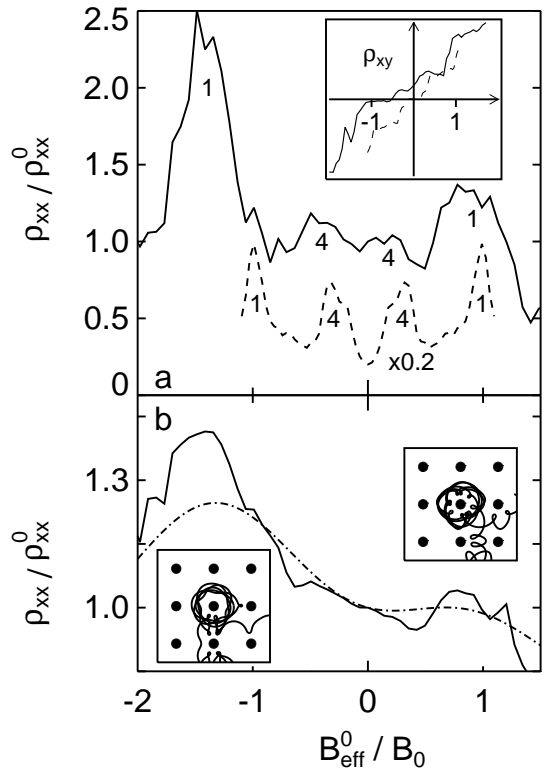

FIG. 1. Results of the numerical integration of the Kubo Formular. (a) Longitudinal resistivity $\rho_{xx}$ of CFs vs. effective magnetic field $B^0_{\text{eff}}$ (solid line) in comparison to $\rho_{xx}$ of electrons vs. magnetic field $B$ (dashed line). Impurity scattering was incorporated in the calculation in relaxation time approximation choosing a mean free path of $\lambda \simeq 25\,\mu m$ in both cases. The magnetic field axis is scaled by $B_0$ corresponding to a free cyclotron diameter equal to the lattice period. For CFs $B_0$ is $\sqrt{2}$ times larger than for electrons due to spin polarization [5]. The resistivities are scaled by the zero magnetic field resistivities $\rho^0_{xx}$. The statistical fluctuations of the data are due to the finite number of orbits used in the numerical calculation. The inset shows the corresponding Hall resistivity of CFs (less $2h/e^2$) which shows shifts in $B$ and $\rho_{xy}$ with respect to the electron result. (b) The longitudinal resistivity $\rho_{xx}$ with mean free path $\lambda \simeq 1\,\mu m$ (solid line) compared to experimental data for lattice period $a = 700\,nm$ by Kang et. al. [4] (dot-dashed line), where the exact position of $B^0_{\text{eff}}$ was determined from the Shubnikov-de Haas oscillations in the vicinity of $\nu = 1/2$. The insets show the different character of chaotic trajectories responsible for the main peaks.

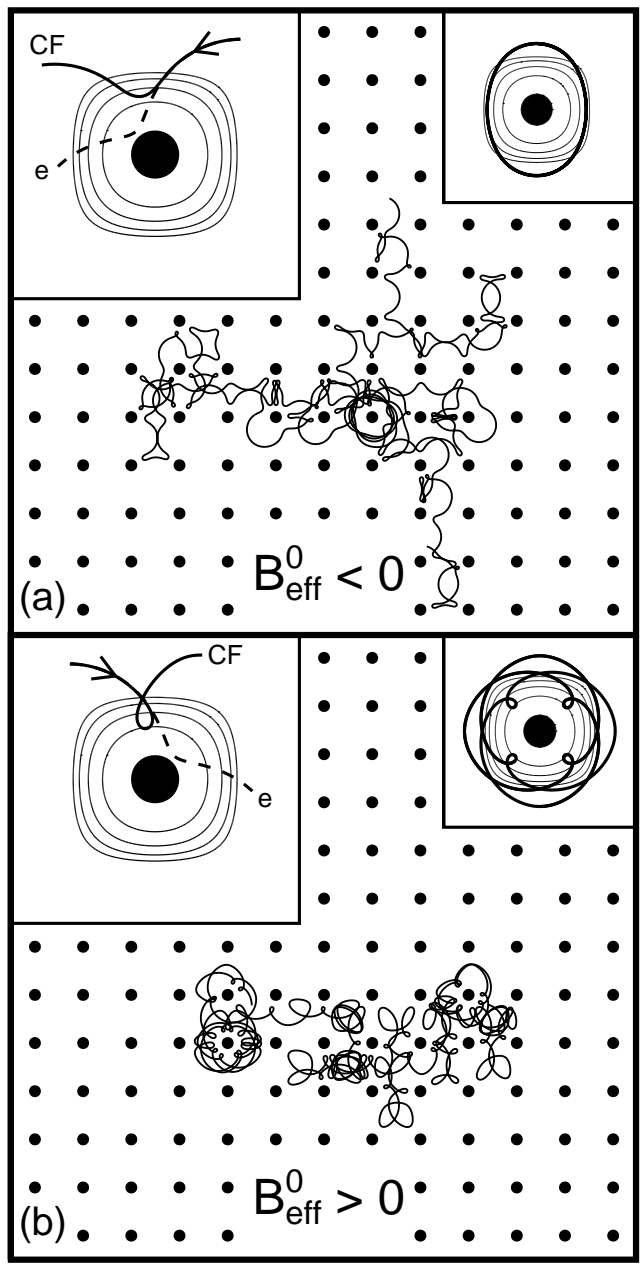

FIG. 2. Segments of chaotic CF trajectories for effective magnetic fields corresponding to the main peak at (a) negative and (b) positive effective magnetic field. The **left insets** show that in contrast to electrons (dashed lines) the CF trajectories (solid lines) are not symmetric under sign reversal of the magnetic field $B^0_{\text{eff}}$, and for $B^0_{\text{eff}} < 0$ their curvature may flip near an antidot. The **right insets** show the different regular orbits belonging to the resonances that trap the chaotic trajectories and thereby are indirectly responsible for the peaks in $\rho_{xx}$. The potential of the antidots is shown as a contour plot, where the black areas correspond to energies larger than $E_F$.

4